# Real-time resilient focusing through a bending multimode fiber

Antonio M. Caravaca-Aguirre,* Eyal Niv, Donald B. Conkey, and Rafael Piestun

*Department of Electrical, Computer, and Energy Engineering, University of Colorado, Boulder, Colorado, 80309, USA*

*Correspondence and request for materials should be addressed to Antonio M. Caravaca-Aguirre, Antonio.Caravaca@Colorado.edu

**Abstract**

We introduce a system capable of focusing light through a multimode fiber in 37ms, one order of magnitude faster than demonstrated in previous reports. As a result, the focus spot can be maintained during significant bending of the fiber, opening numerous opportunities for endoscopic imaging and energy delivery applications. We measure the transmission matrix of the fiber by projecting binary-amplitude computer generated holograms using a digital micromirror device and a field programmable gate array controller. The system shows two orders of magnitude enhancements of the focus spot relative to the background.

## 1. Introduction

In biomedical applications, single mode fibers are normally used for illumination purposes in scanning modalities because of their efficiency in transmitting light[1]. For imaging, single mode fiber bundles are preferred[1]. Unfortunately these may disrupt tissue during insertion because their diameter is on the order of millimeters. As an alternative, if multimode fibers (MMF) could be used to achieve a similar image quality they would become extremely attractive because they possess a smaller cross section and the ability to bend over smaller radii of curvature. However, when light couples into a MMF, scrambling interference among propagating modes is produced because of phase-velocity dispersion and random mode coupling, which change with bending, temperature, and other perturbations; inducing a speckle field at the output of the fiber.

In the 1970's, Yariv et al. provided a theoretical description of image transmission through optical fibers[2,3] and several groups proved the concept later on[4,5]. Lately, there has been a renewed interest in the topic sparked by the new spatial light modulator (SLM), cameras, and computation technical capabilities. For instance, Bianchi et al.[6] demonstrated the generation of multiple spots through a MMF utilizing a liquid crystal (LC) SLM and Monte-Carlo algorithm searching for the optimal input phase mask. In Ref.[7,8] digital optical phase conjugation generates a sharp focus at the output of a MMF which can be scanned for imaging. However, the speed of this system is limited by the refresh rate of the SLM, camera acquisition and processing time. Furthermore, digital holography requires a reference brought through a different channel outside the fiber itself. Other techniques use a transmission matrix (TM) approach allowing bright and dark field imaging[9]. Alternatively, the TM information can be combined with averaged speckle imaging to attain widefield imaging[10]. Random pattern projection followed by optimization of the measurements provides an increase resolution[11]. Importantly, all these techniques rely on a pre-calibration of the optical fiber that typically takes several minutes, after which the fiber should not experience any significant physical change, thus precluding their use in a perturbation prone environment.

In a seemingly different area, two decades ago, Freund introduced the idea of transmitting an image through scattering materials[12]; an idea that only over the last few years has became a reality[13–20]. Vellekoop introduced phase control algorithms using a liquid crystal spatial light modulator (LC-SLM) to overcome the scattering effects produced by turbid media[13,14]. These algorithms are based on a division of the input and output fields into spatial modes, along with the iterative optimization of a merit function. Popoff et al.[15] measured the TM of the scattering material by projecting an orthonormal basis set and measuring the output field for each basis element. After the measurement, the material is characterized and the measured relationship between spatial input and output modes allows focusing to any output mode. A different technique scans different input angles at high speed with the help of galvanometric mirrors[16]. The optimal phase is then found in k-space and transformed to the spatial domain. Moreover, optical phase conjugation[17] using photorefractive crystals and digital optical phase conjugation[18] has been used to compensate for turbidity. Taking advantage of this control, scattering media may be employed as a superlens improving the numerical aperture of an optical system[19]. These techniques not only control the light spatially, but can also compress ultra-short pulses in time[20]. Interestingly, similar techniques have been adopted lately to overcome mode dispersion and coupling in multimode optical fibers[6–11]. The observation is that the speckle field formed at the output of a MMF is similar to the one produced when light propagates through turbid media, even though it is the result of significantly different interfering modes.

Most biomedical applications require placing the fiber within tissue involving shape and temperature changes that produce modifications in the spatial configuration of optical modes. However, it should be emphasized that all these techniques developed for imaging or focusing through MMFs[6–11] are contingent on avoiding any significant disturbance to the fiber during the experiment. Furthermore, a significant issue that limits real time implementation is the latency of the communication between the photodetector, the computer controlling the experiment, and the SLM. Therefore, we address these problems by implementing a real-time phase mask optimization in a MMF setup using a TM algorithm implemented in a field programmable gate array (FPGA).

In Sec. 2 we describe the optical and embedded system, in Sec. 3 the experimental results, and in Sec. 4 we discuss the implications of this work.

**2. Method and Experimental Setup**

*2.1. System*

The setup of the system is shown in Fig. 1. A 532 nm collimated laser beam illuminates a 1024x768 binary amplitude digital micromirror device (DMD TI-DLP Discovery Kit D4100). Each mirror can be controlled to two different angular positions, essentially creating a binary amplitude image. To control the phase of the beam, a binary amplitude Lee hologram[21] encodes the desired wavefront with the DMD. A lens, f1, placed one focal length away from the DMD, Fourier transforms the hologram and an iris in the Fourier plane blocks all the diffraction orders except the -1$^{st}$, which encodes the information of the desired phase distribution. The lens f2 is used to image the phase mask on the back focal plane of a 10X (NA=0.25) objective that couples the light into a 365μm core diameter MMF, 0.22 NA (BFL22-365-Thorlabs), which can propagate an estimated $1.1 \times 10^5$ modes.

At the output of the MMF, the light is received by a 40X (NA=0.65) objective which images the surface of the fiber onto a 50μm pinhole placed before a photodetector. The objective magnification and the pinhole size are chosen to match the pinhole diameter to the speckle spot size at the image plane. To control and characterize the movement of the fiber, we use an automated translation stage that bends the fiber by a measurable angle, θ, shown in the inset of Fig. 1.

A beam splitter placed after the tube lens and before the pinhole creates a second image plane on a CMOS camera (Hamamatsu ORCA Flash 2.8) which enables high frame rate video recording for data analysis and speckle decorrelation time measurement. The photodetector signal is digitized by an analog-to-digital converter (ADC) triggered by the customizable FPGA on the DMD controller board. The intensity values are oversampled and the average value is used to build the hologram with the optimal phase mask encoded to produce the focus spot.

*2.2. Transmission matrix implementation in FPGA*

For real-time experiments, a hardware implementation is often necessary to reduce latency and computation times. A phase shifting method based on three measurements per input mode for TM determination[15,22] is implemented directly on a Virtex5 custom FPGA, APPSFPGA, which is included in the DLP4100 Discovery Kit. A separate analog-to-digital converter (ADC), with a buffered analog input, digitizes the signal and provides the input to the FPGA board. APPSFPGA controls the DMD indirectly via the DDC4100 (digital controller) and DAD2000 (power and reset driver) found on the DLP4100 kit. It is also responsible for accurate triggering of the analog signal conversion as well as storage of the digitized output. For functional verification of the algorithm, modules emulating the DDC4100, DMD and ADC were designed and used during the simulation stage. Figure 2 shows a high-level block diagram of the hardware implementation of the TM focusing system. This configuration is independent of an external computation source and therefore able to toggle the DMD at the maximum frame rate. As a result, the system enables the measurements of the amplitude and phase corresponding to 256 different input modes (with three different reference phases per input mode) in 34 ms. Processing this information to construct the optimal phase mask, then sending to and projecting on the DMD adds an additional 3 ms. It is worthwhile to highlight the speed improvement from our previous report implemented in an application with scattering media[23], where a computer was used and the time spent in this part of the process was in excess of 200 ms. The optimized, focusing phase mask projection time can be varied based on experimental conditions. For this experiment we match focusing time with the measurement time: 37 ms, which provides a 50% duty cycle and allows for repetitive focusing at a constant rate of 13Hz.

## 3. Experiment and results

To quantify the performance of the system, we bend the fiber with different speeds and accelerations to create a dynamic environment. Bending the fiber alters the mode coupling within the fiber; this is manifested by a speckle pattern change at the fiber output. The speckle pattern change is quantified by measuring the 2D correlation between the speckle pattern associated with a bent fiber and a straight fiber. By analyzing the correlation between captured speckle image frames from the CMOS using a static input illumination, the average decorrelation time associated with each setting of the stage is obtained.

To illustrate the sensitivity of the fiber to spatial changes, we measure the focus degradation as the fiber is bent without adaptive correction. In Fig. 3(a) we observe how by displacing the fiber less than 0.1 mm, corresponding to bending the fiber just 0.09º in our configuration, the focus completely disappears. This illustrates the high sensitivity of our fiber to spatial displacements due to mode coupling. To demonstrate the performance of the system, we compare the degradation of the focus spot for two cases: a) with a static phase mask, and b) with adaptive wavefront correction. Figures 3(b) and (c) show the output field intensity of the fiber taken by the camera at different positions of the translation stage. Figure 3(b) illustrates how bending the fiber quickly degrades the created focus when the phase mask is not re-optimized. In Fig. 3(c), the image is shown at the output of the fiber with the adaptive system on. The stage moves with an initial acceleration of 2 mm/s$^2$ until reaching a constant velocity of 1 mm/s, corresponding to an average speckle decorrelation time of 150ms. Despite the fast rate of change, the focus enhancement remains high and stable.

To test the efficiency of the system, we track the dynamics of the focus enhancement as a function of time as the fiber bends, as shown in Fig. 4 (blue line). The zone delimited by the red line corresponds to the interval when the stage is moving. The bending angle of the fiber as a function of time is represented in green. We observe that the enhancement is constant while the stage is not moving. Once the stage starts moving, we observe that the enhancements achieved with the bending fiber are smaller and more variable than with a static fiber, which is due to the dynamics of the transmission matrix; the result of performing measurements while the system is changing.

Furthermore, a higher number of input modes could be used to increase the enhancement as a trade off to speed. Note also that while we present small angle bends because of the range limitation of the automated translation stage, larger angle perturbations have been tested with a manual stage leading to similar enhancement.

**4. Discussion and conclusion**

We have demonstrated real-time focusing through a dynamically bending MMF. In these experiments we used a fiber with more than 10,000 propagating modes to demonstrate the concept. A smaller diameter fiber, which would propagate fewer modes allowing for a larger bending angle before total speckle decorrelation, could also be used. The system creates a focus in 37 ms at the output plane of the fiber leading to enhancements between 50 and 100 during fiber perturbation. As a result of the known relationship between number of modes and focus enhancement, faster operation could be achieved at the expense of enhancement factor. The FPGA implementation enables the system to perform at a constant rate of 13Hz with a 50% duty cycle. The FPGA operation is totally configurable and the phase mask update protocol could be adapted to the application. For example, different algorithms allowing steady state focusing[24] and existing imaging modalities[6–11] could be implemented. The system can have applications in photodynamic therapy where localized energy delivery is needed. A very similar system can be modified for use with dynamic scattering materials, like biological tissue, possessing decorrelation times on the order of ~10 ms[25].

Ongoing research is exploring different feedback mechanisms that can substitute the photodetector at the end of the fiber. For instance, the system could be adapted for imaging purposes by introducing a sparse fluorophore population in the output plane and using the emitted light for feedback[26], or using opto-acoustic effects as demonstrated in scattering media[27].

**Acknowledgments**

We acknowledge support from the National Science Foundation award DGE-0801680. We thank Dr. Miguel Rodriguez and Dr. Carlos Olalla for helpful discussions about the circuit and FPGA design and Yael Niv for helping in the verification process.

**Competing Financial Interests**

The authors declare no competing financial interests

Figure 1

Diagram of the experimental setup. A 532 nm laser beam is encoded with a binary amplitude Lee-hologram displayed on the DMD. The iris placed between lenses f1 and f2 lets through the -1 diffraction order, which carries the encoded information; and is imaged in the back aperture of the objective. The phase mask is focused into the fiber and the output of the fiber is imaged by the f3 lens onto a pinhole in front of a photodetector, whose signal is feed back into the FPGA controller. A CMOS camera images the output plane for monitoring but is not part of the focusing system. The inset shows how we characterize the bending angle. TS: translation stage; BS: Beam splitter; PD: Photodetector; C: CMOS Camera; P: Polarizer; f1, f2, f3: lenses.

Figure 2

Block diagram of hardware implementation. Light propagates through the optical system and comes out of the MMF reaching the photodetector. The signal is digitalized and sent to the DLP4100 kit. The FPGA runs the algorithm with the measured data.

Figure 3

Demonstration of bending resilient focusing: (a) Experimental measurement of the focus degradation as a function of bending angle without active wavefront control. (b) Cross-section of the output intensity of the fiber without running the adaptive wavefront correction, and (c) with adaptive wavefront correction. The focus spot enhancement is shown at the bottom of each cross-section. Scale bar indicates 8μm.

Figure 4.

Enhancement of the focus as a function of time . The red lines delimit the period during which the fiber is being bent. The angle of bending is shown in green. v: velocity of the stage, a: initial acceleration of the stage. $T_c$: Average decorrelation time of the setting.

Figure 1

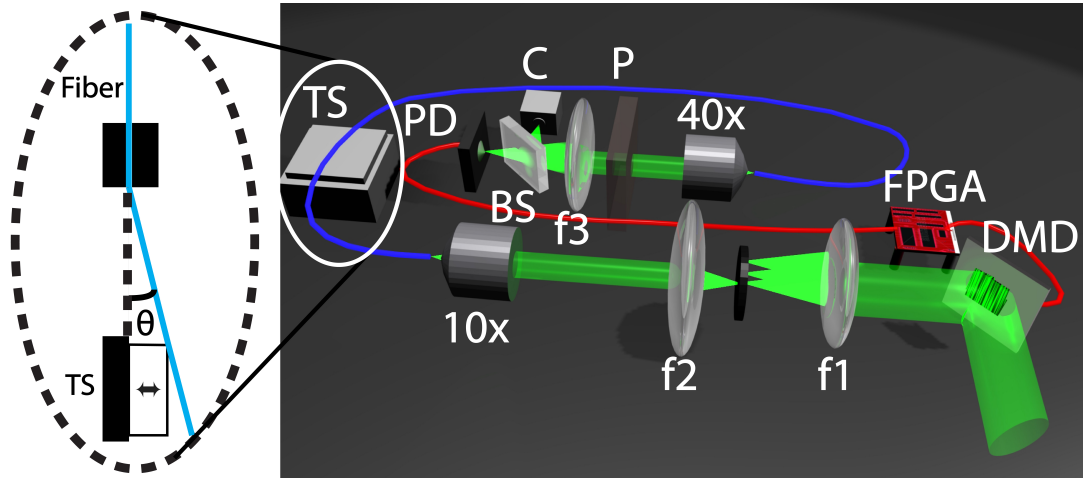

Figure 2

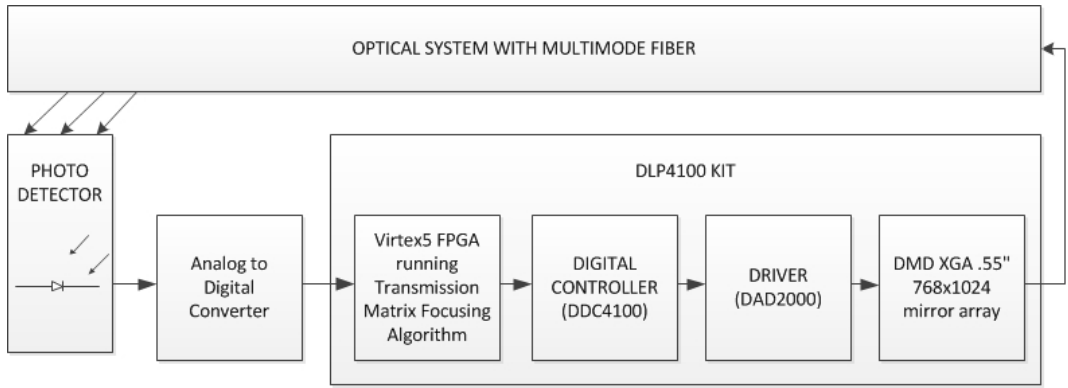

Figure 3

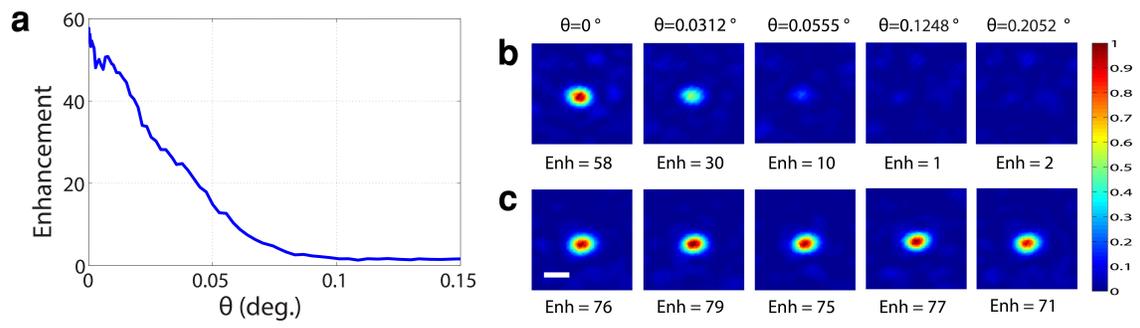

Figure 4

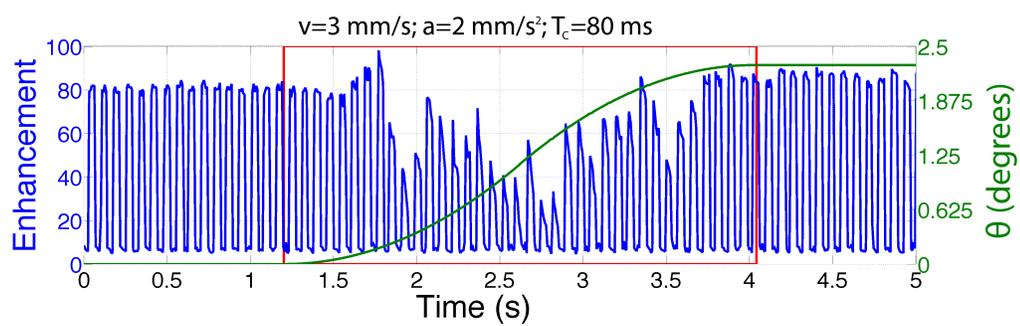